\documentclass[12pt]{iopart}

\usepackage{amssymb,amsbsy,epsfig,color,graphicx}
\usepackage[ansinew]{inputenc}
\usepackage[english]{babel}
\usepackage{braket}
\usepackage{ulem}

\newcommand{\virg}[1]{``#1''}
\newcommand{\bea}{\begin{eqnarray}}
\newcommand{\eea}{\end{eqnarray}}
\newcommand{\be}{\begin{equation}}
\newcommand{\ee}{\end{equation}}

\def\tr{\mbox{tr}\,}

\newcommand{\eu}{{\rm e}}
\newcommand{\ii}{{\rm i}}

\begin{document}

\title{The Frustration of being Odd: Universal Area Law violation in local systems}

\author{Salvatore Marco Giampaolo$\dagger$, Flavia Br\'aga Ramos$\ddagger$ and Fabio Franchini$\dagger$}

\address{$\dagger$ Division of Theoretical Physics, Ru\dj{}er Bo\v{s}kovi\'{c} Institute, Bijeni\u{c}ka cesta 54, 10000 Zagreb, Croatia}

\address{$\ddagger$ International Institute of Physics, Universidade Federal do Rio Grande do Norte, Campus Universit\'{a}rio, Lagoa Nova, Natal-RN 59078-970, Brazil.}

\ead{fabio@irb.hr}

\begin{abstract}
At the core of every frustrated system, one can identify the existence of frustrated rings that are usually interpreted in terms of single--particle physics.
We check this point of view through a careful analysis of the entanglement entropy of both models that admit an exact single--particle decomposition of their Hilbert space due to integrability and those for which the latter is supposed to hold only as a low energy approximation.
In particular, we study generic spin chains made by an odd number of sites with short-range antiferromagnetic interactions and periodic boundary conditions, thus characterized by a {\it weak }, i.e. nonextensive, frustration. 
While for distances of the order of the correlation length the phenomenology of these chains is similar to that of the non-frustrated cases, we find that correlation functions involving a number of sites scaling like the system size follow different rules.
We quantify the long-range correlations through the von Neumann entanglement entropy, finding that indeed it violates the area law, while not diverging with the system size. 
This behavior is well fitted by a universal law that we derive from the conjectured single--particle picture.
\end{abstract}



\maketitle

\section{Introduction}

It is common knowledge that boundary conditions should be chosen wisely when performing numerical simulations, in order not to interfere with the physical phenomenon one wishes to investigate.
On the contrary, in the classification of phases, boundary conditions are supposed to be irrelevant. 
The reason for this apparent paradox is that in the latter case one chooses to take the thermodynamic limit first so that any length scale at which one can probe the system can be considered as \virg{local}, while in the former the finite size of the system inevitably introduces another relevant scale in the game.

However, needless to say, infinite size systems are just an ideal approximation and thus it is important to understand the influence of boundary conditions on finite--size effects, either to avoid them or to exploit them. 
In particular, one question is whether finite size effect decay exponentially or algebraically, since in the latter case the thermodynamic limit has to be treated carefully.
In particular, we will show that this is the case for quantum spin chains with frustrated boundary conditions. 

In general, frustration is the result of competing interactions so that not all terms in the Hamiltonian can be minimized simultaneously.
In this sense, any genuine quantum Hamiltonian includes some amount of frustration, since non--commuting terms promote contrasting local arrangements~\cite{Wolf03,Giampaolo11,Marzolino13}.
However, with the term frustration, one usually refers to the so-called ``{\it geometrical frustration}'', which emerged first in classical systems~\cite{Toulouse77,Vannimenus77}.
Prototypical are models characterized by antiferromagnetic (AFM) interactions with closed loops of odd lengths and every system displaying geometrical frustration, can be explained in terms of the presence of such loops.
In quantum frustrated systems, geometrical and quantum frustration are in general intertwined and it is not easy to discriminate between the two sources~\cite{Giampaolo15}.

To provide an example, the easiest model useful to visualize (classical) geometrical frustration is made by three spins arranged on the vertexes of a triangle, with AFM couplings along with the bonds.
In a classical system with Ising variables as magnetic moments, all interactions cannot be minimized simultaneously, resulting in a \mbox{six--fold} degenerate ground--state. 
It is easy to generalize these considerations for longer spin loops with nearest--neighbor AFM bonds: while on even chains the two N\'{e}el states minimize all local interactions (and thus the whole Hamiltonian), for loops of odd lengths $N=2M+1$, one bond avoid minimization, resulting into a $2N$ degenerate ground--state. 
Promoting the magnetic moments from Ising variables to three--dimensional spins does not alleviate the frustration still resulting in a ground--state degeneracy scaling like the system length~\cite{Sadoc07,Lacroix11,Diep13}. 
It is worth noticing that adding a single site to an AFM loop changes the system dramatically, turning a double degeneracy into a massive one and vice--versa, thus demonstrating that the effect of frustration is {\it non-perturbative} in nature.

In this work, we concentrate on systems with weak, i.e. non--extensive, frustration, such as those of the examples mentioned above, but with the addition of quantum interactions that break their perfect symmetry, thus lifting the degeneracy. 
The traditional expectation, based on a perturbative picture, is that frustrated boundary conditions result in single--particle physics, that is, that ground--state of these systems can be characterized as a single particle excitation over the non--frustrated GS.
Using a combination of analytical and numerical approaches, we check this expectation beyond the perturbative regime.
Consistently with the aforementioned picture, we find that this weak frustration closes the energy gap of a traditionally gapped phase and leads to the appearance of a band of massless excitations with a quadratic spectrum and unusual long--range correlations.
Moreover, we quantify the amount of these long--range correlations, using the entanglement entropy (EE), which is a measure of the entanglement between a portion and the rest of the system.
 
Nowadays the analysis of the EE of the ground--state of a system has emerged as a fundamental probe in the study of quantum complex systems~\cite{Amico08,Calabrese09,Herdman17}, for its ability to detect phase transitions and to characterize phases even beyond the Landau paradigm~\cite{Kitaev06,Levin06}. 
The EE typically follows some universal behaviors for sufficiently large subsystems: while for high energy states it is proportional to the volume of the subsystem, for ground--states of systems with local interactions it satisfies an {\it area} law, with possible logarithmic violations for critical phases~\cite{Vidal03}. 
Intuitively, the area law stems from the fact that entanglement reflects the correlations shared between the subsystems and the rest of the system and these are localized, for gapped systems, in a shell of the order of few correlation lengths around the boundaries. 

In absence of frustration, in one--dimensional models with gapped energy spectrum, the existence of an area--law implies that the EE saturates to a constant value as soon as the dimension of the portion becomes greater than some correlation lengths~\cite{Vidal03,Its05,Peschel04,Franchini07}.
On the contrary, when the energy spectrum is gapless, correlations extend with an algebraic decay,
and thus the EE of the ground--state of one--dimensional systems show the characteristic universal behavior $S(R) \simeq \frac{c}{6} \log R$ of conformal field theories (CFTs) with central charge $c$~\cite{Calabrese04}.

However, the presence of frustration alters this picture.
By performing a careful and in some sense innovative finite--size scaling analysis, in the weakly frustrated case we observe a peculiar violation of the area law, which yet does not result into its divergence for large subsystems, due to its saturation at subsystem lengths proportional to the total system size. 
We quantitatively characterize the observed behavior as due to the contribution, over the non-frustrated GS EE, of a single delocalized excitation, which, therefore, does not possess any intrinsic lengths scale, except for the total system size.

\section{Weakly Frustrated Spin Chains}

Let us introduce a generic nearest-neighbor one-dimensional spin-$\frac{1}{2}$ spin chain with $N$ spins in a magnetic field:
\bea
\!\!\!\!\!\!\!\!\!\!\!\!\!\!\!\!\!\!\!\!H & = & J \sum_{l=1}^{N-1} \left[
 \left( \frac{1 + \gamma}{2} \right) \sigma_l^x \sigma_{l+1}^x +
 \left( \frac{1 - \gamma}{2} \right) \sigma_l^y \sigma_{l+1}^y \right]+J\Delta \sum_{l=1}^{N-1}  \sigma_l^z \sigma_{l+1}^z - \sum_{l=1}^N h \; \sigma_l^z 
\nonumber \\
\!\!\!\!\!\!\!\!\!\!\!\!\!\!\!\!\!\!\!\!&+&  J_N \left[
\left( \frac{1 + \gamma}{2} \right) \sigma_N^x \sigma_{1}^x +
\left( \frac{1 - \gamma}{2} \right) \sigma_N^y \sigma_{1}^y 
\right] + J_N \Delta \sigma_N^z \sigma_{1}^z 
\label{spinham}
\eea
where $\sigma_l^{\alpha}$, with $\alpha=x,y,z$, are Pauli matrices which describe spin-$1/2$ operators on the $l$-th site of the chain.
The Hamiltonian in eq~.(\ref{spinham}) can describe several models with different properties and boundary conditions. 
Choosing $\gamma=0$ we can recover the $XXZ$ model in external fields that holds a continuous $U(1)$ symmetry while for $\gamma \ne 0$ we fall into an $XYZ$ model in an external field characterized by a $\mathbb{Z}_2$ discrete symmetry. The Hamiltonian (\ref{spinham}) is analytically solvable if one of the parameters $\gamma$, $\Delta$, or $h$ is zero.
Settings $J_N = J$ restores translational invariance and choosing $J=1$ (up to an energy scale) favors AFM order that,  
on an odd periodic lattice $N=2M+1$, shows both classical and quantum frustration. 
In this case, the presence of quantum frustration can be proven by settings $h=0$ and observing that the system does not satisfy the quantum Toulouse conditions~\cite{Giampaolo11,Marzolino13}, which discriminates between geometrically and non-geometrically frustrated systems.

The effect of frustration induced by the boundary conditions has been already considered in integrable systems with a continuous $U(1)$ symmetry at vanishing external field as the $XXZ$ chain obtained by setting $\gamma=0$ and $h=0$ in (\ref{spinham})~\cite{Bonner64,Barwinkel03,Cador05,Schnack06}, where the eigenstates can be constructed in terms of individual excitations. 
Thus, while for even lengths $N=2M$ the ground--state can achieve zero total magnetization $S^Z_T = 0$ and be characterized as a spinon vacuum\cite{Franchini17}, in the frustrated case $N=2M+1$ there are two equivalent ground--states with $S^Z_T = \pm \frac{1}{2}$ (whose degeneracy is immediately lifted for a nonzero $h$), which can be interpreted as due to the presence of a traveling single spinon excitation.

The goal of the present paper is to analyze the case of systems with discrete global symmetries $\mathbb{Z}_2$, which, thus, do not conserve particle number.
In Ref.~\cite{Campostrini15}, Campostrini {\it et al.} considered the odd length, ferromagnetic Ising chain, obtained by settings $J=-1$, $\gamma=1$, and $\Delta=0$ in (\ref{spinham}). 
When the defect $J_N$ differs from $J$, it breaks translational invariance and for $J_N>0$ favors AFM order along the $x$-direction between the first and last spins of the chain. 
By varying $J_N$, they found that, for $|h|<1$, $J_N=1$ represents a critical point separating two different phases for $J_N \lessgtr 1$.
Notice that their critical model obtained settings $J_N=1$  can be mapped into the translational invariant AFM Ising chain using local rotations on the even spin sites. 
The authors connect this critical behavior to the metastability of this model under the perturbation provided by a longitudinal magnetic field $\delta H = h_x \sum_{l=1}^N \sigma_l^x$.  
Indeed, it is known that the point $h_x=0$ corresponds to a first--order phase transition~\cite{Campostrini15,Sachdev11}.

The algebraic decay of the correlation functions at $J_N=1$ derived in Ref.~\cite{Campostrini15} was reexamined in Ref.~\cite{Dong16} where Dong {\it et al.} focused on the translational invariant version of the same model. 
In this way, the defect is not localized at the ``end'' of the chain, but it is rather a frustration due to an AFM loop of odd length.
It was observed that this weak frustration is sufficient to scramble the energy spectrum. 
For $|h|<J$, the ground--state is unique with a band of $2N-1$ levels above it, forming a gapless continuum in the thermodynamic limit. 

In their work Dong {\it et al.} notice that the ground state for $|h|<1$ is characterized by two different families of correlation functions.
The difference between these two families can be traced to the different representations of the correlators in terms of spinless fermions. 
Indeed, in one--dimensional system, one can exactly map spins into fermions through the (non--local) Jordan--Wigner transformation~\cite{Jordan28}. 
As a consequence of such non--locality it may happen that the support in which a spin correlator is defined does not coincide with the one of the associated fermionic operators.
To give an example, a spin correlator with support on a finite non--connected region made by two disjoint subsets, can be mapped into a fermionic operator with the support that includes also all sites between the two disjoint subsets. 
We define the correlators whose support in the associated spin and fermionic representation coincide as ``{\it local correlations}'', while the others are the ``{\it quasi--local correlations}''. 

The two families of correlation functions show different behaviors since quasi--local spin correlators present a peculiar algebraic behavior, which is absent for the local ones.
These differences are exemplified by two of the simplest two--point spin correlation functions: the correlation function along $z$ at distance $R$, i.e. $C^{zz} (R)\!\! \equiv\!\! \langle \sigma^z_l \sigma^z_{l+R} \rangle$ and the correlation function along $x$ at distance $R$ i.e. $C^{xx} (R)\!\! \equiv\!\! \langle \sigma^x_l \sigma^x_{l+R} \rangle$.
As it is known in the literature, see for example Ref.~\cite{Barouch71}, while the first is, in agreement with our definition, a local correlation function, the second is a quasi--local one. 
We can easily evaluate the asymptotic behaviors of these two correlators, as a direct extension of the results of~\cite{Dong16}, in combination with the analysis of~\cite{Barouch71}:
\bea
\hskip-1.3cm C^{xx} (R) \label{rhox} &\equiv& \langle \sigma^x_l \sigma^x_{l+R} \rangle 
=(-1)^R \: m_x^2 \left[ 1  + \frac{1}{2 \pi R^2} \frac{h^2}{J^2-h^2} 
\left( \frac{h^2}{J^2}\right)^R \right] \left(1 - \frac{2 R}{N} \right)
 \\
\hskip-1.3cm C^{zz} (R) &\equiv& \langle \sigma^z_l \sigma^z_{l+R} \rangle \label{rhoz}
=m_z^2- \frac{1}{8 \pi} \left( \frac{h^2}{J^2} \right)^{R+1} \!\!\!
+ \frac{4 }{N}  \left[ m_z - \frac{1}{\sqrt{8 \pi}} \left(-\left| \frac{h}{J}\right|\right)^{R+1} \right] 
\eea
where $m_x \equiv \left( 1 - \frac{h^2}{J^2} \right)^\frac{1}{4}$ and  $m_z\equiv \int_0^\pi \frac{h - J \cos \phi}{\sqrt{h^2+J^2-2 h J \cos \phi}} \: \frac{d \phi}{2 \pi} $ are the magnetizations along the two axes. 

To unveil the difference between the two correlation functions in eq.~(\ref{rhox}) and eq.~(\ref{rhoz}) let us consider the values that can be obtained in the thermodynamic limit, i.e. when $N\rightarrow \infty$ ,for spins at very large distances, hence considering $R \rightarrow \infty$.
The limit can be done in two different ways.
We can, at first, place ourselves in a thermodynamically large system (i.e. considering $N\rightarrow \infty$ first) and, only later increase the distance between the two spins. 
Or, on the contrary, at finite $N$ we can set $R$ at the antipodal point, i.e. $R=(N-1)/2$, and then we can make the system size grow.

Using the first approach, taking at first the thermodynamic limit $N\to \infty$, both these two functions reduce to the standard ones of the Ising unfrustrated chain~\cite{Barouch71}, which decay exponentially to saturation, respectively $(-1)^R \! \left(\! 1\! -\! \frac{h^2}{J^2}\! \right)^{\!\!\frac{1}{4}}$ and $m_z^2$,  with correlation length $\xi = - \frac{1}{\ln (h/J)^2}$. 
On the contrary if we first evaluate eq.~(\ref{rhox}) and eq.~(\ref{rhoz}) at antipodal points ($R=(N-1)/2$) and then perform the limit $N \to \infty$, we find that, while nothing changes for $C^{zz}(R)$, whose limit is always $m_z^2$, for $C^{xx}(R)$ we obtain $\lim_{N\to \infty}C^{xx} \left( \frac{N-1}{2} \right)=0$ because of the slow algebraic decay in eq.~(\ref{rhox}).
Hence, differently from the case of eq.~(\ref{rhoz}), the results that we obtain for the limit of $C^{xx} \left(R\right)$ depends on the order in which the two limits are taken. 
This result, is not limited to $C^{xx} \left(R\right)$ but extends to all quasi--local correlation functions that, in the unfrustrated case admit a limit for $R,N\rightarrow\infty$ different from zero.

The unusual behavior of $C^{xx} \left(R\right)$ not only represents a piece of relevant evidence by itself but it also acquires a key role when we take into account that, in the thermodynamic limit, the absolute value of $C^{xx}(R\rightarrow\infty)$ represents the square of the order parameter\cite{Barouch71}.
Exact analytical diagonalization (see supplementary material or \cite{Dong16}) shows that, while without frustration the gap between the ground--state and the first excited state (characterized by opposite parities) closes exponentially in the system size, with frustrated boundary conditions the gap vanishes only polynomially, similarly to the gaps with the higher states. 
Therefore, with frustrated boundary conditions, the asymptotic double degeneracy of the ground--state is missing~\cite{Dong16} and, accordingly, the order parameter should vanish.
This is a surprising result since a nonvanishing longitudinal magnetization is the hallmark of the $\mathbb{Z}_2$ spontaneous symmetry breaking, for which the Ising model is the poster--child~\cite{Barouch71}.

Thus, while locally (i.e. for $R\ll N$) the correlation functions of the frustrated AFM Ising chain are indistinguishable from those of the unfrustrated version, at large distances important differences emerge. 
To capture this diversity one has to consider a {\it scaling thermodynamic limit}, in which distances are measured in terms of the chain length: $r \equiv \frac{R}{N}$, which is kept fixed as $N \to \infty$. 
This limit is equivalent to taking the thermodynamic limit while simultaneously scaling the lattice spacing down as $1/N$. 
Under this scaling limit, {\it quasi-local} correlation functions such as (\ref{rhox}) are characterized by an algebraic decay, as if $\xi\propto N=\infty$.

\section{The Entanglement Entropy}

\begin{figure}[t]
	\begin{center}
		\includegraphics[width=12.cm]{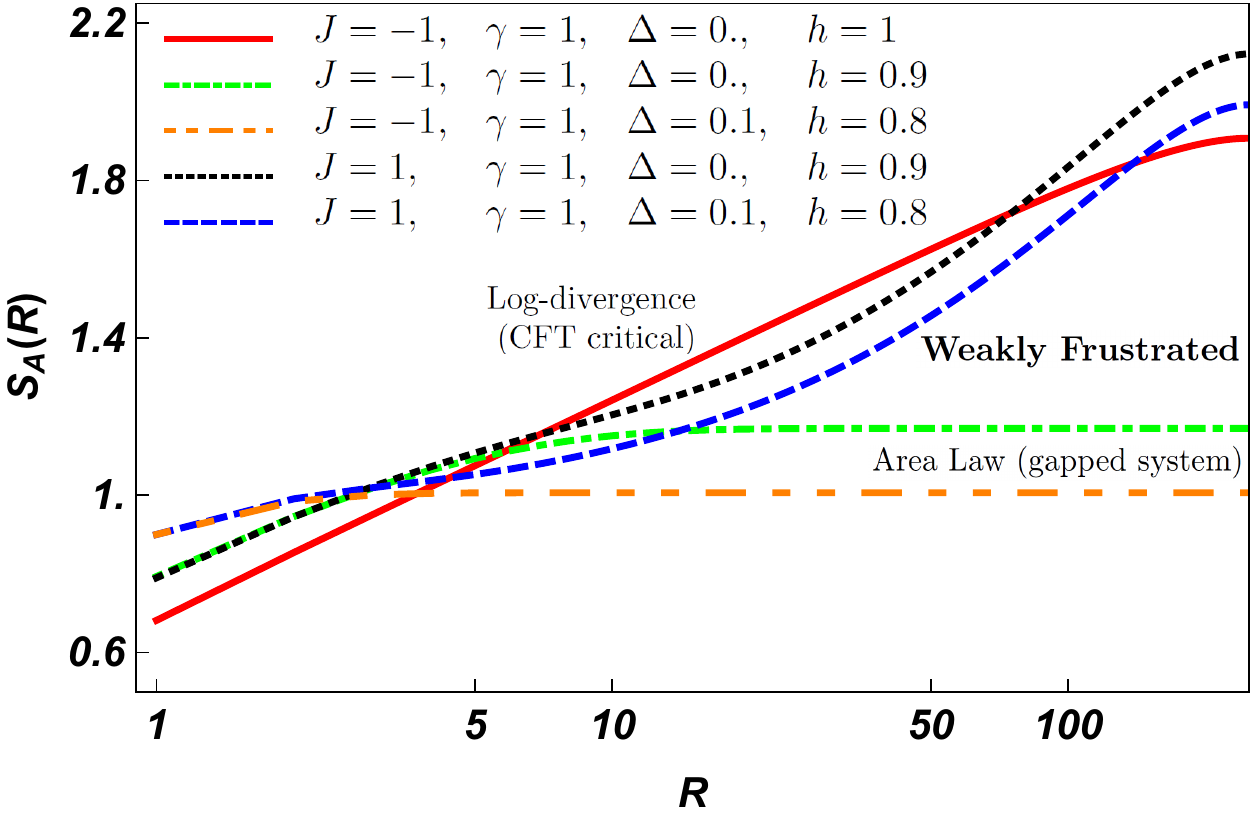}
		\caption{Comparison between the EE of standard phases (gapped and CFT critical) and that of the weakly frustrated case, showing the distinct different behavior of the latter with a violation of the area law. 
        The EE $S_A(R)$ for the reduced density matrix evaluated on a block of $R$ adjacent spins is plotted as a function of $R$ for total chain length $N=501$ and different sets of Hamiltonian eq.~(\ref{spinham}) parameters. 
        The blue and the orange lines are obtained with a numerical DMRG algorithm while all the other lines are obtained semi-analytically using the Jordan--Wigner transformations.
        In considering finite-size systems, it is customary to plot the entropy as a function of $x \equiv \frac{N}{\pi} \sin \frac{\pi R}{N}$, to account for the periodic boundary condition and the symmetry of the entropy around its maximum at $R=2/N$, but here we prefer to show the raw data.}
		\label{EEComparison}
	\end{center}
\end{figure}

To better understand the effects of the frustrated boundary conditions on the ground--states of the Hamiltonian in eq.~(\ref{spinham}) and the emergence of long--range correlations, we focus on the behavior of the EE.
To evaluate the EE, we divide the system into two parts: a subsystem $A$ consisting of $R$ contiguous sites and its complement $B$ with $N-R$ spins. 
We extract the reduced density matrix $\rho_A (R) = \tr_{N-R} |GS\rangle \langle GS|$ of subsystem $A$ and we measure the entanglement between $A$ and $B$ using the Von Neumann entropy~\cite{Bennet96,Nielsen10}, defined as
\be
   S_A (R) = - \tr_A \left[ \rho_A (R) \log \rho_A (R) \right] \; .
\ee

Our analysis will be focused on the characterization of the frustration effects of models with a global discrete $\mathbb{Z}_2$ interactions. 
Hence, in the present paper, we will not take into account, if not marginally, the models that can be obtained setting $h=0$, holding additional $\mathbb{Z}_2$ symmetries, or $\gamma=0$ that show a continuous $U(1)$ symmetry. 
We also limit our analysis to models that are invariant under spatial translation and hence, from now on, we set $J_N=J$.

As we mentioned, our Hamiltonian includes both analytically solvable and non--solvable models. 
Furthermore, for $\Delta=0$ the spin chain is amenable to an exact, although highly non-local, mapping to a free model.
In this case the values of the entropies used in the paper are obtained exploiting the analytical approach based on Jordan--Wigner transformations that is depicted in some details in the supplementary materials (this approach reduces the exponentially complex problem of calculating the EE to the numerical diagonalization of a matrix whose entries are determined analytically and whose rank scales just linearly with the subsystem size). 
On the other hand, for $\Delta\neq 0$ the results for the entropies are obtained using a DMRG algorithm~\cite{White92}.
In the numerical computations, we have considered up to 300 kept states to represent the truncated Hilbert space of each DMRG block. 
Typically, the truncation error is smaller than $10^{-12}$.

\begin{figure}[b]
	\begin{center}
		\hspace{-1.2cm}\includegraphics[width=14.cm]{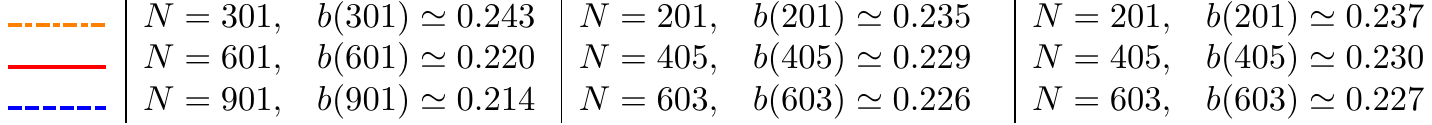}
		\includegraphics[width=14.cm]{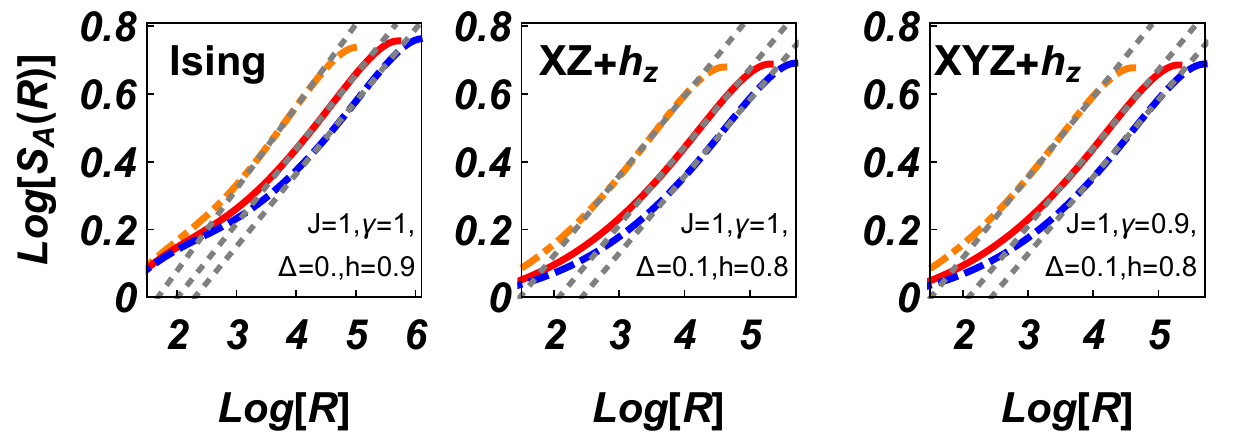}
		\caption{Are--law violation in the weakly frustrated chains. The dependence of the $S_A(R)$ on $N$ is plotted in log--log plot to show that in the bulk it follows a power-law of the type $S_A (R) \simeq a(N) R^{b(N)}$, shown as a dashed gray line. The data of the plot on the left are obtained semi-analytically using the Jordan--Wigner transformations while, in the other two plots, data are obtained with a numerical DMRG algorithm}
		\label{AreaViolation}
	\end{center}
\end{figure}

\begin{figure}
	\begin{center}
		\hskip 1cm\includegraphics[width=7cm]{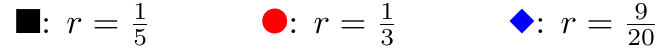}	
		\includegraphics[width=11.6cm]{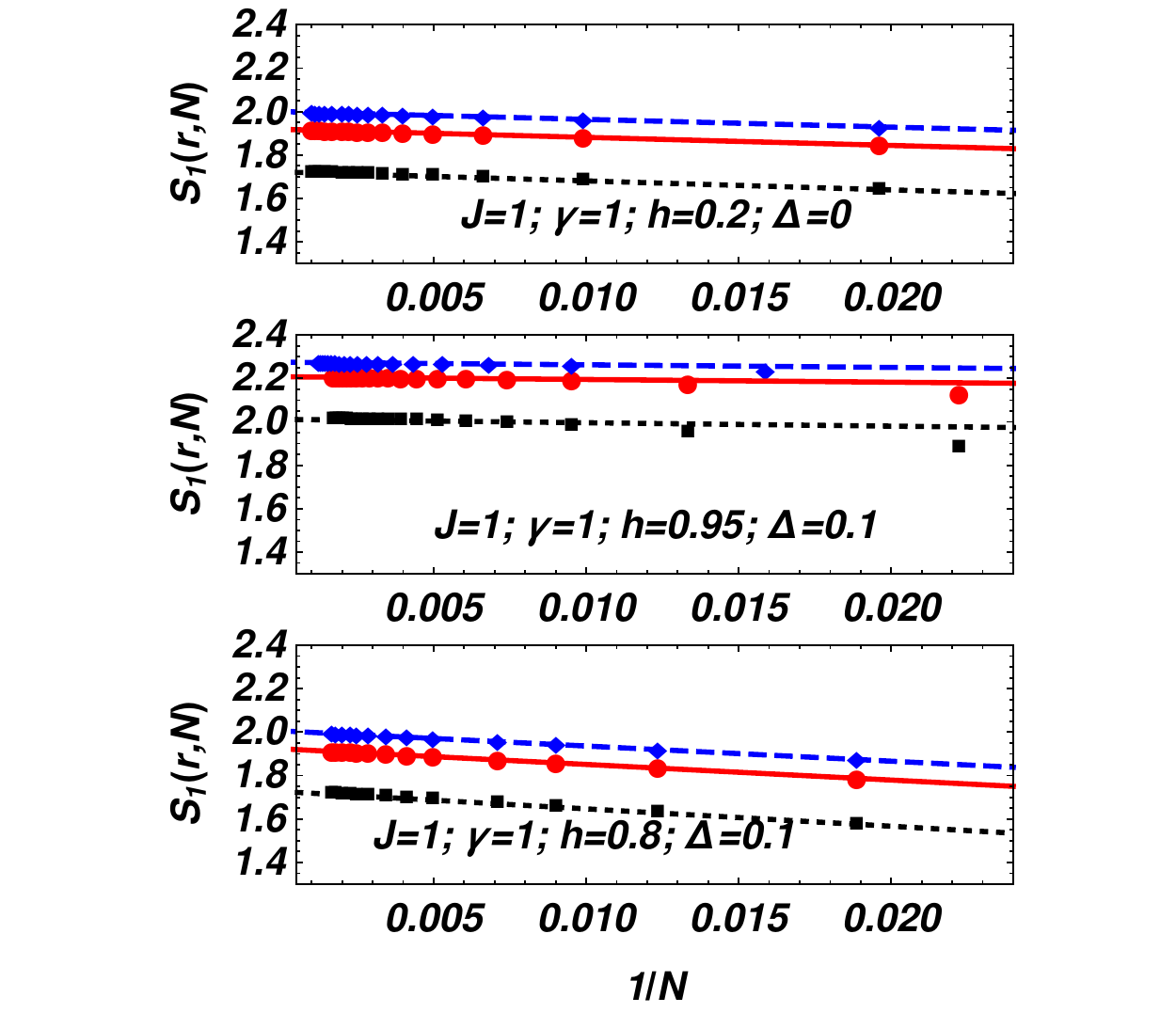}
		\caption{Dependence of the EE $S_A(R)$ on $N$ while keeping the ratio $r=R/N$ constant, in the weakly frustrated chain, for different Hamiltonian parameters. 
        The data of the upper plot are obtained semi-analytically using the Jordan--Wigner transformations while, in the other two plots, data are obtained with a numerical DMRG algorithm.}
        The points represent the values of the entropy obtained, while the lines stand for the best fit with a function of the form $a_r +\frac{b_r}{N}$.
		\label{EESaturation}
	\end{center}
\end{figure}

As we mentioned, the frustrated Ising chain for $|h|< J$ is gapless: this fact and the algebraic decay of some correlation functions point against an area-law behavior.
On the other hand, the spectrum of low energy excitations is quadratic (Galilean) and thus violates relativistic invariance of CFT and hence we have no reason to expect the presence of a logarithmic divergence of the EE~\cite{Brandao13}.
In Fig.~\ref{EEComparison} we observe the peculiar behavior of the frustrated case, compared with the area-law saturation of the corresponding unfrustrated system and the logarithmic divergence at CFT criticality:
\begin{enumerate}
\item For small $R$, compared to the correlation length of the correspondent ferromagnetic model, (i.e. the model obtained changing $J$ in eq.~(\ref{spinham}) from $1$ to $-1$), the EE of the ferromagnetic and the antiferromagnetic systems almost coincide.
\item Increasing $R$ in the unfrustrated case the EE saturates quickly while the frustrated chains still show a growth which is well fitted, in the bulk, by an empirical $S_A (R) \simeq a(N) R^{b(N)}$ where the fitting parameters depend on $N$ as well as on the Hamiltonian ones (Fig.~\ref{AreaViolation}). 
Such dependence on $N$ prevents the EE to diverge in the thermodynamic limit.
\item The saturation of the  EEs in the limit of large $N$ can be appreciated in Fig.~\ref{EESaturation}. 
In the spirit of the scaling thermodynamic limit introduced before, we keep the size of the subsystem $A$ equal to a fixed ratio $r = R/N$ of the total length and plot the EE as $N$ is increased. 
We observe an EE behavior of the type $S_A (N) \simeq a_r + \frac{b_r}{N}$, indicating that in the thermodynamic limit the EE tends to a finite, constant value.
\end{enumerate}

In all plots, we collected data from different points in the phase-space of the generic AFM spin system eq.~(\ref{spinham}), including the Ising chain, the $XY$-chain in a longitudinal magnetic field, and the $XYZ$-chain in an external magnetic field. 
While the Ising chain is akin to a free model, the last two are not even integrable. 
The qualitatively similar behaviors in all these different models are evident.

\begin{figure}
	\begin{center}
		\includegraphics[width=\columnwidth]{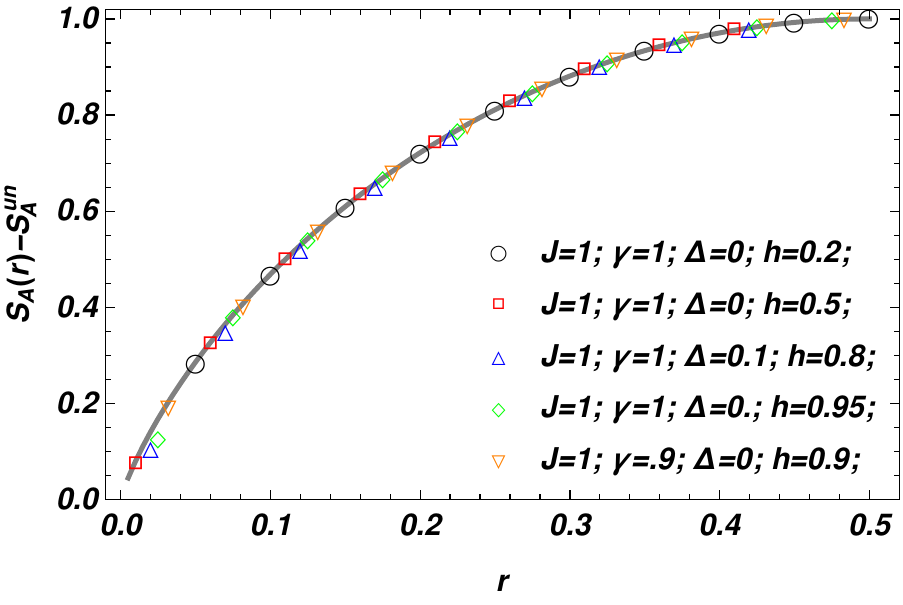}
		\caption{Universal behavior of $S_A(R/N)$ for the weakly frustrated systems in the {\it scaling thermodynamic limit}, once the non-universal, non-frustrated saturation value $S_A^{un}$ of the EE is subtracted.	All numerical data points, extracted in the $N\to\infty$ limit of the EE at fixed $r=R/N$ for different values of the Hamiltonian parameters, perfectly fall on a single line, plotted in gray, given by eq.~(\ref{universal}).}
		\label{Collapse}
	\end{center}
\end{figure}

This agreement can also be made quantitative. 
Collecting all entropy saturation points in the $N \to \infty$ limit for the different values of the parameters in the same plot, we observe in Fig.~\ref{Collapse}, that they all fall on the same {\it universal curve}, once the non-universal, non-frustrated saturation value is subtracted. 
This is quite surprising because previous studies of models with a Galilean invariant spectrum have either given different behaviors~\cite{He17,Gentle18} or very non-universal ones~\cite{Ercolessi11,CastroAlvaredo11,CastroAlvaredo12,Alba12}. 

We can fit this universal curve using the single--particle picture of the frustrated ground state. For instance, at $\gamma\!=\!\Delta\!=\!h\!=\!0$, the ground--state of the frustrated system can be interpreted as a superposition of domain walls. Turning on slightly any of the above parameters introduces some hopping so that the ground--state can be approximated as a traveling excitation. 
Thus, the entanglement entropy can be estimated to be $\log 2=1$ from the double degeneracy of the Neel states plus a contribution due to the probability that the domain wall excitation lies or not in the interval $A$:
\be
	\rho_A (R) = \frac{R}{N} \ket{1}\bra{1} 
	+ \left( 1 - \frac{R}{N} \right) \ket{0}\bra{0} \; ,
	\label{rhoA01}
\ee
where $\ket{0},\,\ket{1}$ indicates a state with the excitation inside/outside of the subsystem (note that eq.~(\ref{rhoA01}) is valid also for non--point--like excitations, as long as translational invariance is assumed).
A more refined approach to be applied further away from the$\gamma\!=\!\Delta\!=\!h\!=\!0$ point includes the fact that the non--frustrated ground--state possess a structure and a finite, non--trivial entanglement:
\be
	\ket{GS_{unfrustrated}} = \sum_\alpha \sqrt{\lambda_\alpha} \ket{\psi_\alpha^A} \ket{\psi_\alpha^B} \; ,
\ee
where we employed the usual Schmidt decomposition of a state~ \cite{Nielsen10}. In the single--particle interpretation, the reduced density matrix, in this case, is constructed as
\be
	\rho_A (R) = 
	\sum_\alpha \lambda_\alpha \: r \ket{\psi_\alpha^A,1}\bra{\psi_\alpha^A,1}
	+ \sum_\alpha \lambda_\alpha \left( 1 - r \right)  \ket{\psi_\alpha^A,0}\bra{\psi_\alpha^A,0}\; ,
	\label{rhoARGS}
\ee
and its EE is
\bea
	S_A (r) & = & - \sum_\alpha \lambda_\alpha \log \lambda_\alpha 
	- r \log r - (1-r)\log(1-r) 
	\nonumber \\
	& = & S_A^{un} - r \log r - (1-r)\log(1-r) \; ,
	\label{universal}
\eea
where the first, constant terms, is the non-universal saturation value of the EE for the un-frustrated case.
This is the curve plotted in Fig.~\ref{Collapse} and it is in excellent agreement with the numerical data, indicating that, even for non-integrable models which lack an exact quasi-particle description, the EE of very large systems does not show deviations from a single particle approximation.

It should also be noted that eq.~\ref{universal} 
differs from
\be
  S_A(r) \simeq 
  S_A^{un} + \sqrt{8} \Big( r (1 -r) \Big)^{\frac{3}{4}}
  \label{universal2} 
\ee
at most by 0.7\% and hence the two curves are virtually indistinguishable from each other. The latter expression of the entanglement entropy explains the  algebraic area law violation which we already noticed for finite systems in Fig.~\ref{AreaViolation}. 
Notice that eq.~(\ref{universal2}) is a simple power-law only for small values of $r=R/N$, while trying a power-law fit for larger values of $r$ results in a varying exponent, thus explaining the fit in Fig.~\ref{AreaViolation}.

In conclusions, the results in Fig.~\ref{Collapse} are in strong contrast both with the divergence shown by standard (CFT) critical models and with the exponential convergence to a constant value that is found in systems satisfying the area law.
This behavior is consistent with that of a single excitation on top of a non-frustrated ground state: while the latter is characterized by a finite correlation length which sets its saturation, the former does not have any intrinsic lengths scale, except for the total system size. We observe that this picture is quite general and robust and not related to specific, fine-tuned models.

\section{Discussion and Conclusions}

We have shown how a weak (nonextensive) frustration induced by the boundary conditions can deeply affect the properties of generic quantum spin chains whose Hamiltonian holds a $\mathbb{Z}_2$ discrete symmetry, with the appearance of a mixture of correlation functions with exponential and algebraic decay.
The latter is very slow, since the relevant parameter is $r=\frac{R}{N}$, and arise as a consequence of the non-trivial boundary conditions. 
We characterized this emerging pseudo-phase using the EE: it shows a violation of the area law with an algebraic growth with the subsystem, which yet does not lead to divergence for large systems. 
Such behavior supports the idea that, as in gapped chains, the total amount of entanglement in the system is finite, but, similarly to critical systems, correlations are distributed through the whole chain, with the possibility of distilling Bell-pairs with arbitrary distance~\cite{Bennet96,Bennet96_1,Popp05}.

Frustrated boundary conditions are often considered to result in a single particle excitation. 
Accordingly, the EE is interpreted as due to the superposition of a ground--state contribution (characterized by a finite correlation length) and a delocalized excitation (with infinite correlation length). 
We calculated the EE within such a picture in eq.~(\ref{universal}) and confirmed in Fig.~\ref{Collapse} the quantitative agreement between the analytical expression and the numerical data for a variety of frustrated spin chains.

Indeed, at least for the Ising chain, the ground--state of the frustrated chain has the same correlation functions of certain low--lying states of the non--frustrated case.
It should also be remarked that, in system lacking particle number conservation such as the one we have analyzed, and with no integrability to characterize states in terms of quasi-particle excitations, the ground--state selected by the boundary conditions does not present any simple exact characterization, but our entanglement data shows that it is consistent with a single--particle interpretation.

Frustrated boundary conditions are a way to render otherwise low energy states stable against decay, with possible application for state engineering for quantum technologies.
Moreover, the considerations above imply that low energy states (of non--frustrated models) carry much more structure than previously noticed, with very long--range correlations (scaling like the system size) which could be harvested for quantum information processing or transmission and quantum criptography~\cite{Lo99}.
As we mentioned, these states seem to have a finite amount of entanglement but spread peculiarly. 
And it is known that, for several tasks, it is not important the total amount of entanglement in a system, but how it is distributed~\cite{Pan01}.
We plan to investigate these perspectives in our next works. 
For instance, preliminary results show that the phase diagram of the frustrated pseudo--phase is quite rich and includes regions with degenerate ground--states with peculiar properties, such as the spontaneous breaking of translational invariance. 

Although to the best of our knowledge, the EE behavior we observed has not been reported in any system before, this is not the first class of local, translational invariant systems which presents a violation of the area law. 
Recently, two such examples have been introduced, i.e. the Motzkin~\cite{Bravyi12} and the Fredkin chains~\cite{Dellanna16}. 
These are frustration-free systems, in the sense that the Hamiltonian can be decomposed as a sum of local commuting terms, all sharing the same ground--states. 
This feature also allows for a direct evaluation of their entanglement entropy, which scales either logarithmically with the subsystem size for low--spin chain, or as a square--root for higher spins--variable lengths. 
These models share similarities and profound differences with the class of weakly frustrated systems we considered. 
For instance, both are related to a {\it massive degeneracy} of the ground--state manifold, but in a very different way. 
For such systems, a massive degeneracy exists for periodic boundary conditions, but the area law violation requires an open chain with certain conditions at the borders, which selects from the manifold a unique, highly entangled, ground--state. 
In the frustrated case, the massive degeneracy is lifted by the external magnetic field and periodic boundary conditions are crucially needed to enforce frustration and observe the area law violation. 
Also, in the frustration--free models, the area law violation is accompanied by a divergence of the EE for large systems, which is not the case for the weakly frustrated cases.  
Most of all, the frustration--free systems are somewhat artificial in their construction, especially so for the cases of square-root violation of the area law. 
On the contrary, the frustrated systems we considered are very natural and robust against perturbations.

Although we considered only 1D chains with weak frustration, we remark that these are at the core of any frustrated system, even in higher dimensions, where frustration is always produced by closed loops~\cite{Toulouse77,Vannimenus77}. A certain degree of frustration is very common and can give rise to peculiar properties: systems with an extensive amount of frustration (i.e. a number of loops proportional to the size of the system), both regular, such as the ANNNI model~\cite{Fisher80} or spin ices~\cite{Bramwell01}, and disordered , such as the Sherrington-Kirkpatrick model~\cite{Sherrington75} and spin glasses~\cite{Binder86}, showcase unique behaviors different from those of unfrustrated systems, such as algebraic decay of correlation functions without criticality~\cite{Huse03,Henley05}, local zero-modes~\cite{Villain79,Ritchey93,Oleg2002,Lee02}, residual entropy at near-zero temperature~\cite{Harris97,Ramirez99}, and give rise to peculiar emergent properties, such as artificial electromagnetism~\cite{Huse03,Henley05} monopoles, and Dirac strings~\cite{Morris09}. 
Also, magnetic frustrated systems are among the best candidate to host the elusive spin liquid phase~\cite{Balents10}.

An important outcome of our work is that even weakly frustrated systems can present peculiar behaviors if observed at a length scale comparable to the loop size.
We can thus speculate that some of the properties of strongly frustrated systems (which have loops of many different lengths) have their origin in the phenomenology we discussed in this work.
We plan to address this hypothesis by considering extensively frustrated quantum chains, to characterize the resulting phase using the scaling thermodynamic limit we introduced. 
This analysis would be an important step toward the consideration of generic frustrated systems. 
As closed loops are the building blocks for general frustrated systems, embedding the considerations we developed in higher--dimensional systems can help to better understand the interplay between geometrical frustration and quantum interaction and to decipher the complicated behaviors of frustrated systems.

\section*{acknowledgments}
We thank Andrea Trombettoni, Rosario Fazio, Marcello Dalmonte and Alexander Abanov for useful discussions and Ryan Requist for his careful reading of the first version of the manuscript and his comments. We are grateful for the computational resources provided by the High Performance Computing Center (NPAD) at UFRN.  
FF and SMG acknowledge support from the H2020 CSA Twinning project No. 692194, ``RBI-T-WINNING'' and from RBI TWIN SIN project. 
FF's work is also partially suppported by the Croatian Science Fund Project No. IP-2016-6-3347.

\section{Supplementary Material: The Weakly Frustrated Ising Chain}

Although our results for the generic $XYZ$ chain show that the weakly frustrated pseudo-phase is quite general and not limited to the odd AFM Ising chains, it is instructive to look in details at the antiferromagnetic Ising model to see how these unusual behaviors emerge.

Let us specialize (\ref{spinham}) to $J=J_N=1$, $\gamma=1$, and $\Delta=0$:
\be
H_{\rm Ising} = 
\sum_{l=1}^{N} \left(
\sigma_l^x \sigma_{l+1}^x - h \; \sigma_l^z \right) \; ,
\label{Isingham}
\ee
where periodic boundary conditions $\sigma^\alpha_{l+N}=\sigma^\alpha_l$ are assumed. 
The nearest neighbor interactions and the external magnetic field are non-commuting terms thus providing a quantum nature to the model.
The $\mathbb{Z}_2$ symmetry of the model is implemented by the parity operator $\mathbb{P}=\prod_{l=1}^N \sigma_l^z$. Such operator measures the parity of the magnetization along the $z$-axis,
admits two degenerate eigenvalues $P=\pm 1$  and commutes with the Hamiltonian $\left[ H_{\rm Ising}, \mathbb{P} \right]=0$.

To study this chain, the standard procedure is to first apply the Jordan-Wigner transformation (JWT) which maps spin-$1/2$ variables into spinless 
fermions\cite{Jordan28}:
\be
\sigma_l^+ =
\eu^{\ii \pi \sum_{j<l} \psi_j^\dagger \psi_j} \; \psi_l \: , 
\qquad \qquad
\sigma_l^z = 1 - 2 \psi_l^\dagger \psi_l \; ,
\label{JordanWigner} 
\ee
so that an empty fermionic site corresponds to a spin up, with further phase decoration due the non-local string in (\ref{JordanWigner}). 
Although the JWT solves the difficult problem of dealing with spins, it explicitly breaks translational invariance, by selecting a first site from which the string starts. 
Because of this, the Hamiltonian written in terms of fermions presents a defect, set by the parity operator $\mathbb{P}$, in the coupling between the first and last spin. 
One way to deal with this issue is to separate from the beginning the Hilbert space into the two subspaces of different parities. 
Then, the defect is removed by imposing periodic or antiperiodic boundary conditions to the fermionic system depending on the parity, which, in turn, is reflected in the choice of integer/half-integer quantization for the Fourier momenta. 
Finally, the Hamiltonian in Fourier space is quadratic and can be diagonalized exploiting a Bogoliubov rotation~\cite{Franchini17}. 
After this sequence of non-local mapping, the Ising chain (\ref{Isingham}) is transformed exactly into the free fermionic Hamiltonian
\bea
H &=& \frac{1 + \mathbb{P}}{2} \; H^+ \; + \; \frac{1 - \mathbb{P}}{2} \; H^- \; , 
\nonumber \\
H^\pm &=& \sum_{q \in \Gamma_\pm}
\varepsilon \left( \textstyle{ \frac{2 \pi}{N} \: q } \right) \;
\left\{ \chi_q^\dagger \chi_q - \frac{1}{2} \right\} \; ,
\label{H+ham}
\eea 
with spectrum
\be
\varepsilon (\alpha) \equiv \sqrt{ (h + \cos \alpha)^2 + \; \sin^2 \alpha},
\label{spectrum}
\ee
with the exception of momenta $\frac{2 \pi}{N} q=0,\pi$, since these modes have energy $h \pm 1$ respectively.
The set of allowed momenta depends on the parity and is given by $\Gamma_\mathbb{P}=\left\{n+ \frac{1+P}{4}\right\}_{n=0}^{N-1}$.
The $0$- and $\pi$-modes are special: in the unfrustrated cases, they are responsible for the double degeneracy in the symmetry broken phase~\cite{Franchini17}, while for the frustrated chain they close the gap. Let us discuss only the latter case here.

The absolute ground--state of (\ref{H+ham}) for $N=2M+1$ belongs to the even parity sector and is always the vacuum of Bogoliubov fermions $\chi_q \ket{GS} =0$ for $q=\frac{1}{2}, \ldots,N- \frac{1}{2}$. 
For $|h|<1$ it has energy 
\be
E_0 = - \frac{1}{2}  \; \sum_{q=0}^{2M} \varepsilon \left[
\textstyle{ \frac{2 \pi}{N} \left( q + \frac{1}{2} \right) }
\right] + 1 -h \; .
\label{E0}
\ee 
The $\pi$-mode (corresponding to $q=M$) has negative energy and so its absence costs energy. 
However, it cannot be occupied alone, because such state would have odd parity and does not belong to the same Hilbert space.
Note that in the odd parity sector an exact $\pi$-mode is not allowed because of the (integer) quantization condition for the momenta and thus the odd parity sector does not have a negative energy mode. 
Therefore, the lowest energy excited states in the even parity sector are of the type $\chi^\dagger_{M+1/2} \chi^\dagger_{p+1/2} \ket{GS}$ and have energies 
$E_p = - \frac{1}{2} \; \sum_{q=0}^{2M} \varepsilon \left[\textstyle{ \frac{2 \pi}{N} \left( q + \frac{1}{2} \right) }
\right] + \varepsilon \left[\textstyle{ \frac{2 \pi}{N} \left( p + \frac{1}{2} \right) }\right]$, 
which lie arbitrarily close to $E_0$, with a quadratic dispersion: 
$E(k) \simeq E_0 + \frac{1}{2} \left( \frac{h}{1-h} \right) (k-\pi)^2 + \ldots$. 
In the thermodynamic limit, this set of states form a continuum above the ground--state. 
In the odd parity sector, the lowest energy state has energy greater than $E_0$ and also lies at the bottom of a quadratic gapless band of $N$ states $\chi^\dagger_p \ket{GS'}$ (where $\ket{GS'}$ is the state annihilated by all the $\chi_q$, for $q=0,\ldots,N-1$), where $p=M,M+1$ has the lowest energy. 
As $N \to \infty$, the bands in the even and odd sector mix, with the energy difference between the lowest energy states in the two sectors vanishing polynomially. 
In total, the ground--state is part of a band of doubly -- and in some points four-times-- degenerate $2N$ states.

A special role in this construction is played by the negative-energy mode, whose occupation reduces the total energy of the system.
The crucial difference between the frustrated and the non-frustrated case is that in the former this mode appears in the even parity sector and cannot be occupied alone, while in the latter belongs to the odd parity sector and thus lowers the energy of the lowest energy state, while not closing the gap with the rest of the band~\cite{Franchini17}. 
Also, as we mentioned, the energy difference between the lowest energy states in the two sectors closes polynomially in $N$ in the weakly--frustrated  pseudo--phase and exponentially in the ferromagnetic phase of the non-frustrated models.

One can visualize what happens in the frustrated phase starting from the classical point $h=0$.
In this case, for $N=2M$, the ground--state would be given by one of the two N\'{e}el states. 
However, moving from even to odd $N$, since these states do not satisfy anymore the AFM condition for a pair of neighbor spins, they are degenerate with the additional $2N-2$ states with one domain wall. 
Turning on a finite $h$ splits this degeneracy, but, unlike what happens to other very symmetric points under perturbations, in this case the gap between the states is not proportional just to the strength of the perturbation $h$ and thus these $2N$ state fan out into the band discussed above~\cite{Dong16}.

Having the ground--state representation in the free fermionic language allows for the calculation of the physical spin correlation functions, by inverting the transformations sketched above~\cite{Franchini17}. 
Even more striking, from the fundamental two-point functions one can construct the correlation matrix, whose eigenvalues provide the diagonal form of the reduced density matrix needed for the EE, as explained in~\cite{Vidal03}.
Defining the (Majorana) fermionic operators $ A_l  \equiv \psi_l^\dagger + \psi_l $ and $B_l  \equiv \imath(\psi_l - \psi_l^\dagger)$, both the spin correlation functions and the correlation matrix can be expressed in terms of three kind of expectation values, i.e. $\langle A_l A_m\rangle$, $\langle B_l B_m\rangle$ and $\langle A_l B_m\rangle$. 
The first two of them, for both the frustrated and the unfrustrated Ising model, are $\langle A_l A_m\rangle=\langle B_l B_m\rangle=\delta_{l,m}$. 
The third one, $\langle A_l B_m\rangle$, is non-trivial: we exploit translational invariance to set $l=m+r$ and write $\langle A_l B_m\rangle=\imath G(r,J,h)$ where the $G(r,J,h)$ function satisfies the following properties
\be
\label{rel_corr}
G(r,1,h)=-G(r,-1,-h)+\frac{2}{N} \nu(h,r)
\ee
where $\nu(h,r)$ is equal to $(-1)^r$ for $h>0$ and $-1$ for $h<0$.
We observe that, compared with the unfrustrated case, the presence of a weak frustration adds a weak term to this correlation function that scales as $1/N$.
Even if it seems a negligible contribution, it can play a key role. 

Indeed, since this model is quadratic, all correlation functions can be expressed using Wick theorem in terms of the fundamental two-point functions above.
The spin correlation functions that we call ``{\it local}'' are represented through a finite number, say $K$, of two-point functions. 
Thus, the contributions due to frustration are of the order $K/N$ and vanish in the thermodynamic limit. This is the case of the two-body correlation function along $z$ in (\ref{rhoz}).
On the contrary, the ``{\it non-local}'' spin correlation functions hold an expression, in terms of the fermionic ones, in which the number of terms increases with the distance, typically because of the Jordan-Wigner string in (\ref{JordanWigner}). 
In such cases, the role played by the contribution $\nu(h,r)$ must be taken into account also in the scaling thermodynamic limit and leads to an algebraic decay, as for (\ref{rhox}). 

A fortiori, in agreement with the picture mentioned above, the EE, which can be evaluated in terms of the eigenvalues of the correlation matrix, 
can be considered as a correlator involving a number of two-point functions $ G(r,J,h)$ growing with the subsystem size.
This fact is consistent with the common-sense knowledge that the EE is a non-local quantity. 


\begin{figure}
	\begin{center}
		\includegraphics[width=8.6cm]{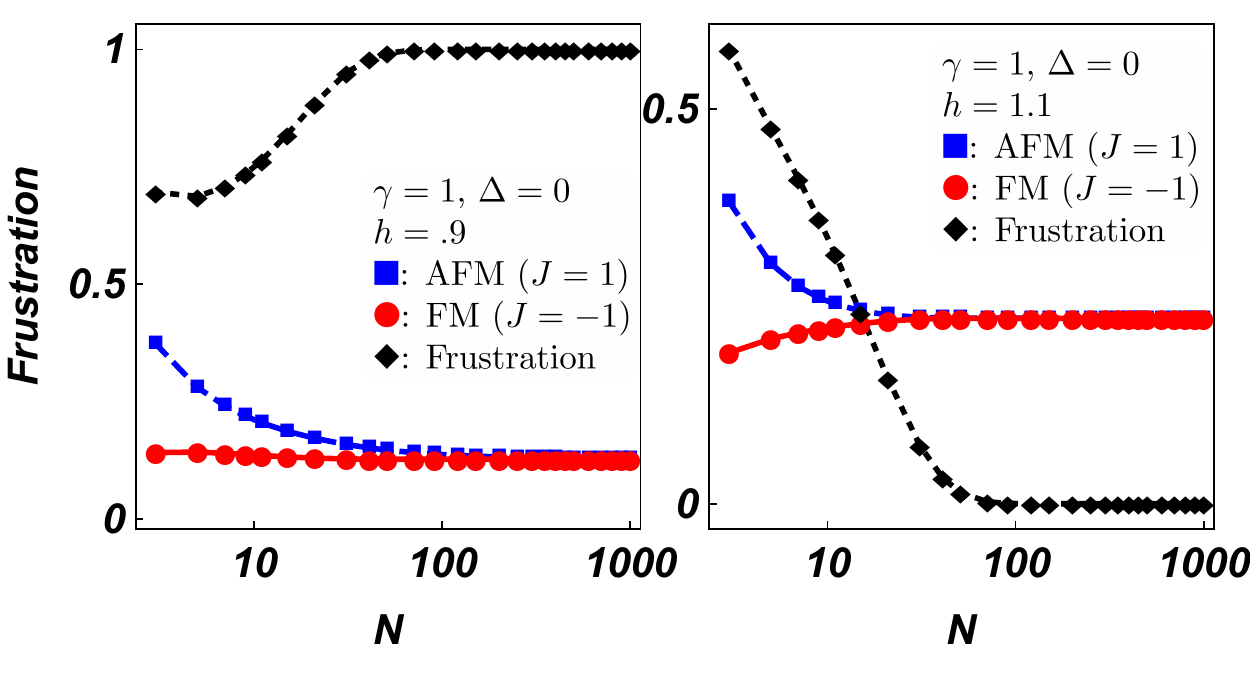}
		\caption{Frustration as a function of the size of the system, for the two phases of the Ising model. Representing the Hamiltonian as a sum over local interactions, the blue(red) points/lines represent the amount of frustration of a single such interaction term in the AFM(Ferromagnetic) case, respectively, while the black curve is their difference summed over the whole chain (\ref{geometric}), representing the amount of geometrical frustration. 
			The {\it quantum} phase $h<1$ is the one which spontaneously break the $\mathbb{Z}_2$ symmetry for $J=-1$ and generates the frustrated pseudo-phase for $J=1$ and is the only one showing a finite amount of frustration (indicating the the single interaction difference scales like $1/N$ in the frustrated phase and $N^{-\alpha}$, with $\alpha>1$, otherwise). Similar results hold for the generic $XYZ$ chain.
		}
		\label{Frustration}
	\end{center}
\end{figure}

To further analyze the role of the weak frustration, we present in Fig.~\ref{Frustration} the behavior of the frustration measure $F(J,\gamma,\Delta,h)$ defined in~\cite{Giampaolo11} for every single interaction. 
As in completely unfrustrated systems each term in the Hamiltonian can be minimized independently, this measure of frustration coincides with the Hilbert-Schmidt distance between the projector in the local ground--space (i.e. the subspace in which every single interaction would take the system if all the other terms of the Hamiltonian were turned off) and the ground--state that is actually realized for the whole system.
As the distance increases, the frustration grows.
Notice that, due to its definition, such a measure of frustration cannot discern between quantum and geometrical frustration. 
Since the ferromagnetic model presents only the former, to distill the contribution of the latter we may use the following quantity:
\be
\label{geometric}
g_F= \sum_{j=1}^N \big[ F(1,\gamma,\Delta,h)-F(-1,\gamma,\Delta,h) \big] \, ,
\ee
where, in fact, the sum is over identical contributions due to translational invariance.
In other words, we estimate the weight of the geometrical frustration as the extra amount of frustration in the antiferromagnetic system with respect to the ferromagnetic case.
As we can observe in Fig.~\ref{Frustration}, for large $N$, while in the paramagnetic phase $g_F$ vanishes, in the new phase it goes to a constant value.
Similar results hold also for $\Delta \neq 0$.
This is in perfect agreement with the na\"ive observation that the amount of geometrical frustration does not increase with the length of the chain.

\appendix

\end{document}